\documentclass[conference]{IEEEtran}
\IEEEoverridecommandlockouts
\usepackage{cite}
\usepackage{amsmath,amssymb,amsfonts}
\usepackage{hyperref}
\usepackage{algorithmic}
\usepackage{graphicx}
\usepackage{textcomp}
\usepackage{booktabs}
\usepackage{xcolor}
\def\BibTeX{{\rm B\kern-.05em{\sc i\kern-.025em b}\kern-.08em
    T\kern-.1667em\lower.7ex\hbox{E}\kern-.125emX}}
\begin{document}

\makeatletter
\newcommand{\rmnum}[1]{\romannumeral #1}
\newcommand{\Rmnum}[1]{\expandafter\@slowromancap\romannumeral #1@}
\makeatother

\title{A Tool for Semantic-Aware Spatial Corpus Construction\\
}

\author{
    \IEEEauthorblockN{Wei Huang}
    \IEEEauthorblockA{
    \textit{College of Computer Science and Technology}\\
    \textit{Nanjing University of Aeronautics and Astronautics}\\
    Nanjing, China\\
    sx2416105@nuaa.edu.cn}\\

    \IEEEauthorblockN{Jianqiu Xu\thanks{\textsuperscript{*} Corresponding author. The authors acknowledge the support of the National Natural Science Foundation under grants Nos.(U23A20296, 62472217).}\textsuperscript{*}}
    \IEEEauthorblockA{
    \textit{College of Computer Science and Technology}\\
    \textit{Nanjing University of Aeronautics and Astronautics}\\
    Nanjing, China\\
    jianqiu@nuaa.edu.cn}
    \and
    \IEEEauthorblockN{Xieyang Wang}
    \IEEEauthorblockA{
    \textit{College of Computer Science and Technology}\\
    \textit{Nanjing University of Aeronautics and Astronautics}\\
    Nanjing, China\\
    xieyang@nuaa.edu.cn}\\
    
    \IEEEauthorblockN{Guidong Zhang} 
    \IEEEauthorblockA{
    \textit{School of Automation}\\
    \textit{Guangdong University of Technology}\\
    Guangdong, China\\
    guidong.zhang@gdut.edu.cn}
}

\maketitle

\begin{abstract}
Spatial natural language interface to database systems provide non-expert users with convenient access to spatial data through natural language queries. However, the scarcity of high-quality spatial natural language query corpora limits the performance of such systems. Existing methods rely on manual knowledge base construction and template-based dynamic generation, which suffer from low construction efficiency and unstable corpus quality. This paper presents semantic-aware spatial corpus construction (SSCC), a tool designed for constructing high-quality spatial natural language query and executable language query pair corpora. SSCC consists of two core modules: (\rmnum{1}) \textit{a knowledge base construction module} based on spatial relations, which extracts and determines spatial relations from datasets, and (\rmnum{2}) \textit{a template-augmented query pair corpus generation module}, which produces query pairs via template matching and parameter substitution. The tool ensures geometric consistency and adherence to spatial logic in the generated spatial relations. Experimental results demonstrate that SSCC achieves (\rmnum{1}) \textit{a 53× efficiency improvement for knowledge base construction} and (\rmnum{2}) \textit{a 2.5× effectiveness improvement for query pair corpus}. SSCC provides high-quality corpus support for spatial natural language interface training, substantially reducing both time and labor costs in corpus construction.

\end{abstract}


\section{Introduction}
Spatial databases serve as fundamental infrastructure for managing geographic data across diverse domains, including location-based services, urban planning, and environmental monitoring \cite{b1}. The SECONDO \cite{b3} spatio-temporal database system provides robust support for spatio-temporal data management, enabling complex spatial queries. However, the development of natural language interface to database (NLIDB) \cite{b4} faces a critical bottleneck due to the scarcity of high-quality spatial natural language query (NLQ) corpor. Unlike relational databases that have abundant labeled NLQ and executable language (NLQ-EXE) query pairs \cite{b5}, spatial databases lack publicly available datasets, making manual annotation both costly and time-consuming.

In the era of large language models (LLMs), solving NLIDB problems requires high-quality corpora, especially for domain-specific databases like spatial databases, which is adopted in a wide range of applications, the demand for high-quality spatial query pairs increases rapidly, while current template-based, rule-based, machine learning, and hybrid approaches struggle to meet this demand. The inherent complexity of spatial relations, coupled with the requirements for semantic accuracy and geometric validity \cite{b6}, presents significant barriers to scalable corpus construction. These challenges can be summarized in three critical limitations \cite{b7} that hinder the development of robust spatial query understanding systems.

\textbf{Scalability and quality inconsistency.} 
Manual annotation methods face severe scalability constraints and inconsistent quality across annotators. The specialized knowledge required for spatial query annotation, including the understanding of geometric operations and spatial relations, makes consistency difficult to maintain. As a result, the quality of manually annotated corpora varies noticeably, rendering large-scale corpus construction impractical and expensive.

\textbf{Inadequate usage of spatial knowledge.} 
Existing automated methods rely on simple template substitution without ensuring the validity of spatial relations. Such approaches often produce semantically invalid or geometrically impossible queries, as the generated spatial relations may not correspond to actual configurations in the database. This lack of geometric determination can mislead models during training and degrade downstream NLIDB performance.

\textbf{Absence of systematic quality control.} 
Most methods lack systematic mechanisms for assessing and refining corpus quality.
Without reliable quality control, errors in the generated data cannot be efficiently identified or corrected, resulting in unstable training datasets that hinder model robustness.

\begin{figure*}[htbp]
\begin{center}
\includegraphics[width=1.0\linewidth]{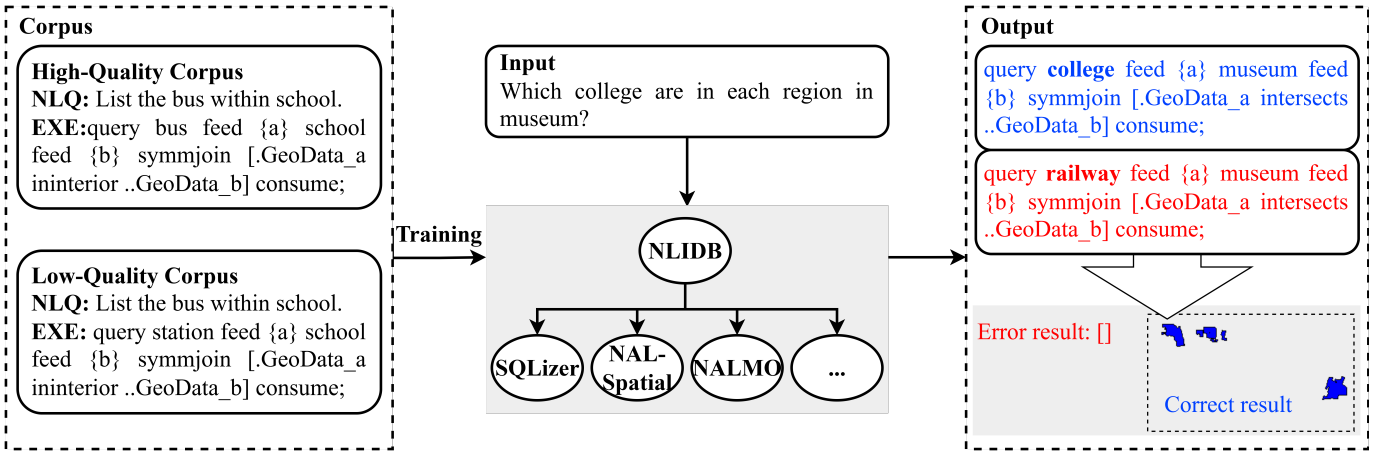}
\end{center}
\caption{An example of corpus quality impact on NLIDB conversion.}
\label{fig1}
\vspace{-1.5em}
\end{figure*}

A deeper analysis reveals that the limitations of existing methods stem from three fundamental challenges: scalability, spatial knowledge validity, and quality control. Current manual annotation methods struggle with scalability due to the specialized knowledge required for spatial query annotation, which makes maintaining consistency across annotators difficult. This results in variable quality in the generated corpora, making large-scale corpus construction impractical. Automated methods, such as template-based approaches, fail to ensure the validity of spatial relations, often generating semantically invalid or geometrically impossible queries that do not align with actual database configurations. Furthermore, many existing methods lack systematic mechanisms for quality control, preventing efficient identification and correction of errors in the generated data, which leads to unreliable datasets and hinders the development of robust spatial query understanding systems (as illustrated in Fig. \ref{fig1}).

To address these challenges, we present semantic-aware spatial corpus construction (SSCC), a tool for the automatic construction of high-quality spatial query pairs.
SSCC employs a two-stage framework comprising two modules:
(\rmnum{1}) \textit{knowledge base construction}, which determines spatial relations through geometric analysis to eliminate invalid entries, and
(\rmnum{2}) \textit{template-augmented query pair generation}, which produces query pairs using template matching and spatial constraints to maintain semantic consistency and geometric validity.

\begin{figure}[htbp]
\begin{center}
\includegraphics[width=1.0\linewidth]{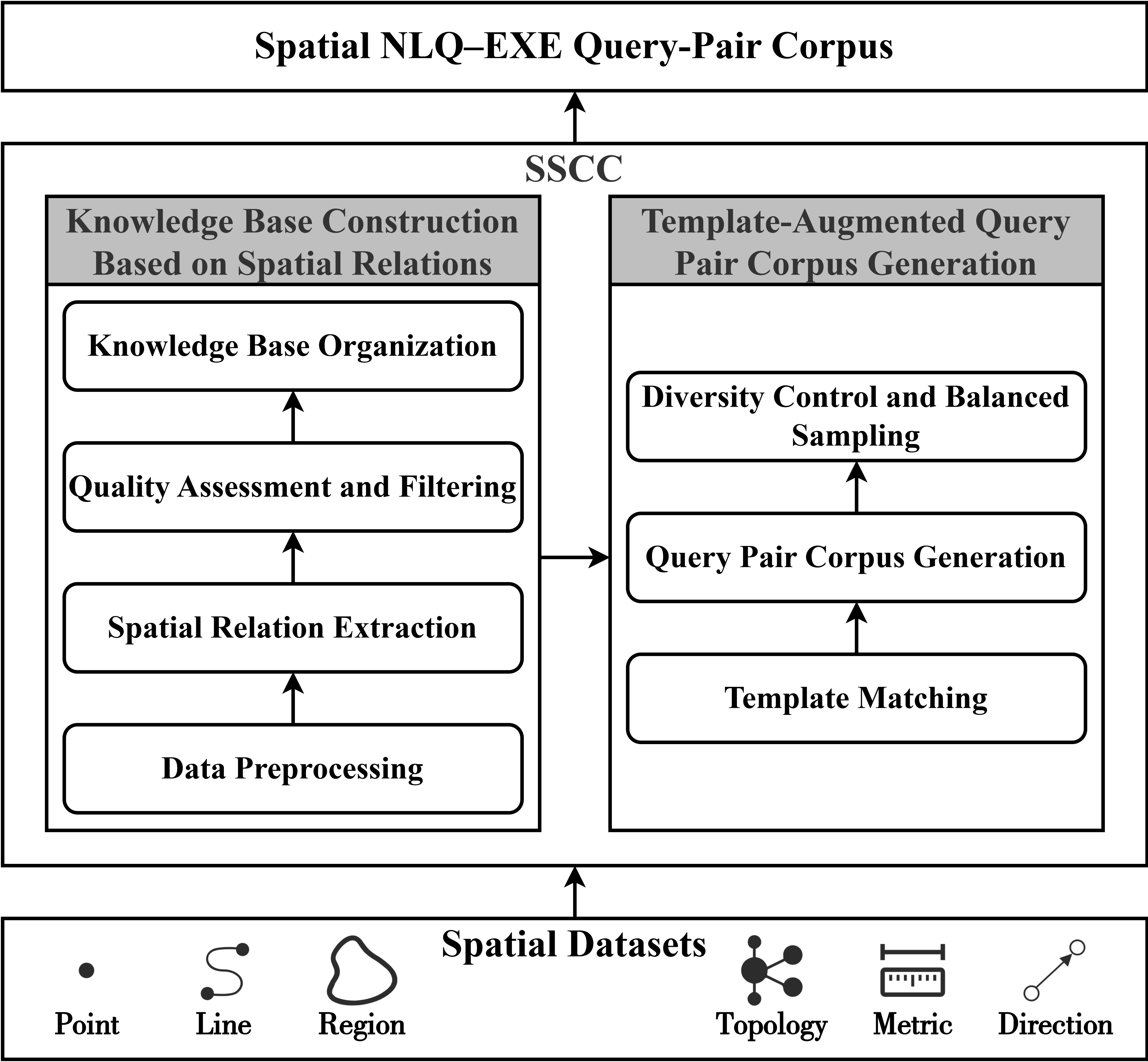}
\end{center}
\caption{Overview of the SSCC framework.}
\label{fig2}
\vspace{-1.5em}
\end{figure}

In this demo, we present the implemented SSCC tool, featuring a Web-based interface and a back-end integrated with the SECONDO spatial database.
To evaluate the tool’s effectiveness, we adopt two key metrics: knowledge base construction throughput and query pair corpus effectiveness, and compare SSCC with SpaCor \cite{b2}, a baseline method employing manual knowledge base construction and template-based dynamic generation.
Experimental results show that SSCC achieves substantial improvements, with (\rmnum{1}) \textit{a 53× increase in knowledge base construction throughput} and (\rmnum{2}) \textit{a 2.5× improvement in corpus effectiveness}.
The knowledge base–driven framework ensures that the generated query pairs maintain high semantic fidelity and geometric soundness, enabling efficient construction of semantically consistent spatial corpora tailored to specific databases.

\section{The Framework}
The SSCC framework aims to construct a high-quality NLQ corpus that is semantically aligned with spatial databases.
As shown in Fig. \ref{fig2}, the tool consists of two main components:
(\rmnum{1}) \textit{knowledge base construction} based on spatial relations, and
(\rmnum{2}) \textit{template-augmented query pair corpus generation}.

\subsection{Knowledge Base Construction Based on Spatial Relations}
Dynamic template filling methods rely on manually extracted entity relations and often lack geometric determination.
To ensure the geometric validity of spatial relations, SSCC computes spatial quality scores to filter relations.

\textbf{Data preprocessing.} 
The tool parses spatial datasets to identify geometric objects such as points, lines, and regions along with their attributes.
Each geometric object is converted into a Shapely geometric structure \cite{b8} for efficient computation.
An STRtree spatial index \cite{b9} is constructed to accelerate geometric operations and relation extraction.

\textbf{Spatial relation extraction.} 
Spatial relations are extracted through a set of geometric operations, divided into two categories:
(\rmnum{1}) \textit{entity–relation queries}, which calculate relations between entities and reference objects \textit{(e.g., distance, intersection, and containment)}, and
(\rmnum{2}) \textit{relation–relation queries}, which analyze relations among different spatial relations \textit{(e.g., spatial joins, distance joins, and aggregations)}.
The tool supports multiple spatial operators, including \textit{intersects, inside, distancescan,} and \textit{symmjoin}, to handle diverse spatial relations.

\textbf{Quality assessment and filtering.} 
The tool evaluates spatial relations using geometry-specific scoring strategies considering distance decay, overlap ratio, and weighted composite scores.
For multi-entity relations, a weighted evaluation mechanism combines average quality, intersection and distance rationality.
Relations exceeding a predefined quality threshold are retained in the knowledge base to ensure spatial reliability.

\textbf{Knowledge base organization.} 
Determined relations are organized into structured knowledge base files consisting of entity–relation tables and relation–relation tables.
The entity–relation tables store spatial relations between entities and reference objects, while the relation–relation tables store relations among spatial relations.
Each record includes attributes such as \textit{entity name, geometry type, distance}, and \textit{operator}, supporting efficient template matching and parameter substitution during query generation.

\subsection{Template-Augmented Query Pair Corpus Generation}
To generate high-quality spatial query pair corpora, SSCC employs a template-driven strategy.
This approach retrieves and instantiates spatial relations from the geometrically determined knowledge base to fill predefined templates, ensuring semantic and geometric consistency in the generated pairs.

\textbf{Template matching.} 
The tool maintains a structured template library covering major spatial query types, including range queries, nearest-neighbor queries, spatial joins, distance joins, and aggregation queries.
Each template contains parameter placeholders and type constraints.
The tool associates templates with corresponding knowledge base entries through a matching algorithm based on query compatibility and parameter constraints, ensuring correctness in parameter substitution.

\textbf{Query pair corpus generation.} 
The tool performs template instantiation by mapping spatial relations from the knowledge base to template parameters.
Entity placeholders are replaced with geographic entities from the knowledge base, distance parameters are automatically converted according to template units, and spatial operators are dynamically selected based on geometry types.
The final query pairs maintain semantic alignment through a parameter synchronization mechanism between natural language and structured queries.

\textbf{Diversity control and balanced sampling.} 
To prevent bias caused by frequent entities, SSCC implements diversity control to limit entity repetition.
A balanced sampling strategy ensures uniform distribution across query types and geometric combinations.
By enforcing sampling constraints and weight allocation, SSCC produces balanced corpora that cover diverse spatial query scenarios, improving representativeness and generalization of training data.

\section{Demomstration}
\begin{figure*}[htbp]
\begin{center}
\includegraphics[width=1.0\linewidth]{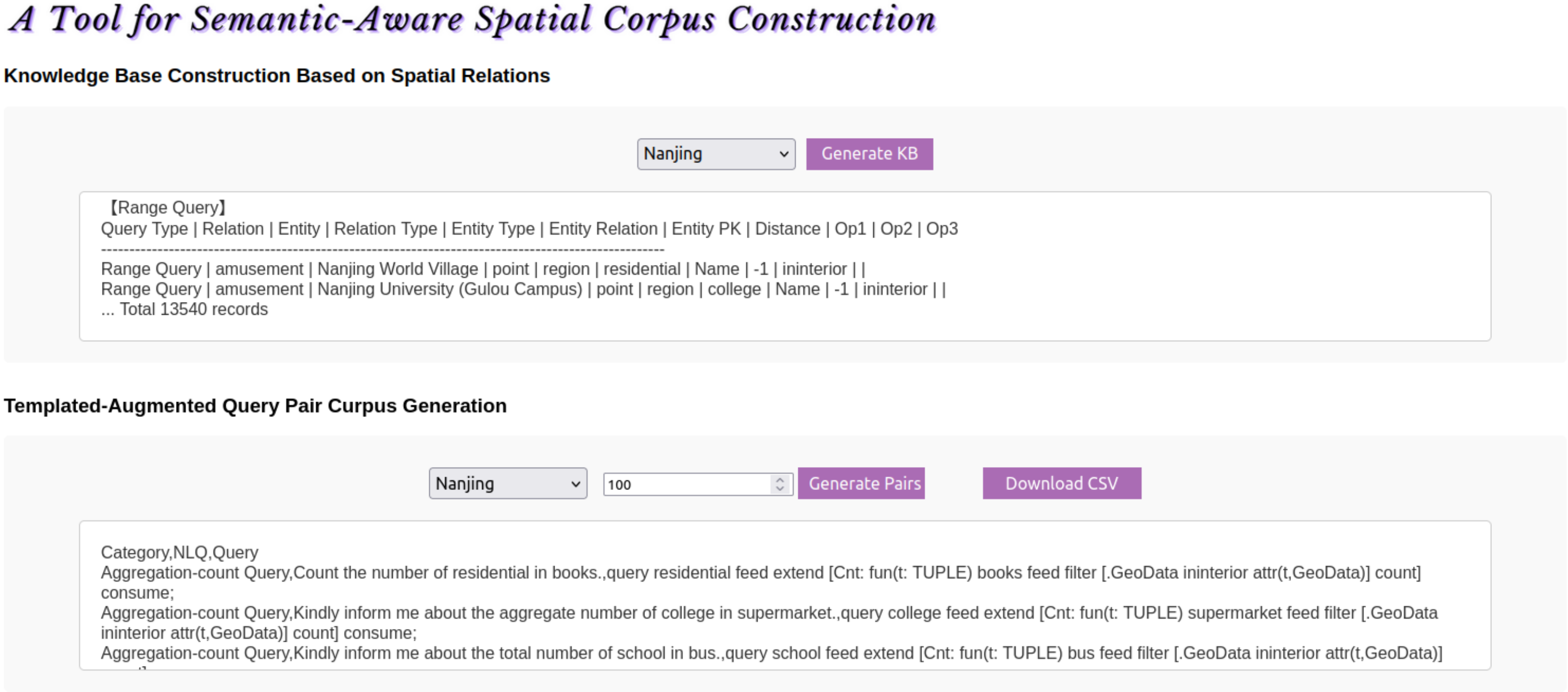}
\end{center}
\caption{ The screenshot of SSCC.}
\label{fig3}
\vspace{-1.5em}
\end{figure*}

We implement the SSCC tool and conduct comprehensive experimental evaluations.
Fig. \ref{fig3} presents the demonstration interface and primary functionalities of SSCC.
The experiments demonstrate two core modules of the tool:
(\rmnum{1}) \textit{knowledge base construction} based on spatial relations, and
(\rmnum{2}) \textit{template-augmented query pair corpus generation}.
Through testing on multiple spatial datasets, SSCC efficiently constructs high-quality spatial query corpora, showing clear improvements over existing baseline methods.

The tool supports multiple spatial datasets, including those of Berlin and Nanjing in Table \ref{tab1}, and automatically executes the knowledge base construction pipeline, which includes geometric object recognition, spatial indexing, relation extraction, and quality filtering.
In the query pair generation module, SSCC generates high-quality query pairs through template matching and parameter substitution.
The Berlin dataset contains 1,898 point features, 4,356 line features, and 261 polygon features, resulting in 6,515 extracted spatial entities.
The Nanjing dataset includes 2,321 point features, 8,407 line features, and 1,942 polygon features, resulting in 12,670 extracted entities.
Each dataset undergoes preprocessing to ensure geometric integrity and spatial accuracy.

The tool is deployed on an Ubuntu 20.04 LTS server with Python 3.8, using SECONDO for spatial data storage, Shapely for geometric computation, and STRtree for spatial indexing.
The tool supports batch processing and real-time updates, enabling efficient handling of large-scale spatial data.

\vspace{-0.5em}
\begin{table}[htbp]
\caption{Experimental Test Dataset Distribution}
\begin{center}
\begin{tabular}{ccccc}
    \toprule
    \textbf{Database} & \textbf{\#Tables} & \textbf{\#Points} & \textbf{\#Lines} & \textbf{\#Regions}\\
    \midrule
    Berlintest & 24 & 1898 & 4356 & 261\\
    Nanjingtest& 18 & 2321 & 8407 & 1942\\
    \bottomrule
\end{tabular}
\label{tab1}
\end{center}
\vspace{-2em}
\end{table}

\subsection{Demonstration Platform Functionality}
\textbf{Knowledge base construction based on spatial relations.} 
Users can select spatial datasets such as Berlin or Nanjing for knowledge base construction.
The tool automatically performs data preprocessing, geometric recognition, spatial indexing, relation extraction, and quality filtering.
Extracted relations are grouped and displayed by query type, including \textit{query type, relation1, relation2, relation1\_type, relation2\_type, distance,} and \textit{operator}.
The tool also provides statistics summarizing entity and relation counts.

\textbf{Template-augmented query pair corpus generation.} 
Users can specify the target dataset and the number of query pairs to generate (default: 100).
The tool performs template matching, parameter substitution, and query generation.
Generated query pairs are displayed in the interface.
A “Download CSV” option is provided for users to export the generated corpus for further analysis and use.

\subsection{Performance and Results Analysis}
To verify the effectiveness and practicality of SSCC, a comparison is made with SpaCor \cite{b2}, a baseline using manual knowledge base construction and template-based generation.
The evaluation includes two aspects:
(\rmnum{1}) \textit{knowledge base construction throughput}, and
(\rmnum{2}) \textit{query pair corpus quality}. 

\textbf{Knowledge base construction throughput.} 
We compare SSCC’s automated knowledge base construction with SpaCor’s manual process.
Experimental results show that SSCC completes the Nanjingtest dataset construction within 3.5 minutes, generating 13,765 valid spatial relations, achieving a throughput of 65.5 items/s.
In contrast, SpaCor requires approximately 3.2 hours to construct a knowledge base of the same scale, with a throughput of only 1.2 items/s.
SSCC improves construction throughput by 53×, leading to substantial reductions in both time and labor costs.

\textbf{Query pair corpus quality assessment.} 
We randomly sample 156 NLQs from SpaCor’s corpus and manually annotate their corresponding executable language to form a benchmark test set.
Then, SSCC generates the same number of query pairs for comparison.
Query pair validity is evaluated based on whether the generated query pairs can successfully execute and return meaningful results in the target database.
Results show that SSCC achieves 91.7\% query pair validity, a 2.5× improvement over SpaCor’s 25.6\%, demonstrating SSCC’s effectiveness in enhancing corpus quality through geometric relation determination and template matching.

As shown in Table \ref{tab2}, SSCC achieves notable improvements in both knowledge base construction efficiency and corpus quality, providing high-quality data support for spatial NLIDB system training.

\vspace{-0.5em}
\begin{table}[htbp]
\caption{Performence Comparison Between SSCC and SpaCor}
\begin{center}
\begin{tabular}{ccc}
    \toprule
    & \textbf{Throughput(items/s)} & \textbf{Corpus Effectiveness(\%)}\\
    \midrule
    SpaCor & 1.2 & 25.6 \\
    \textbf{SSCC} & \textbf{65.5} & \textbf{91.7}\\
    \bottomrule
\end{tabular}
\label{tab2}
\end{center}
\vspace{-2em}
\end{table}

\vspace{12pt}

\end{document}